\documentstyle[prl,aps,graphics,epsfig,floats]{revtex}
\begin{document}
\draft
\twocolumn
\title{Observation of  power-law scaling for phase
transitions
in linear trapped ion crystals}
\author{D. G. Enzer, M. M. Schauer, J. J. Gomez, M. S. Gulley, M. H.
Holzscheiter,  P. G. Kwiat, S. K. Lamoreaux, C. G. Peterson, V. D.
Sandberg,  D. Tupa, A. G. White, R. J. Hughes}
\address{Physics Division, Los Alamos National Laboratory, Los Alamos,
NM 87545, USA} 
\author{D. F. V. James}
\address{Theoretical Division, Los Alamos National Laboratory, Los
Alamos, NM 87545, USA} 
\date{\today}
\maketitle

\begin{abstract}
We report an experimental confirmation of the power-law relationship between
the critical anisotropy parameter and ion number for the linear-to-zigzag phase
transition in an ionic crystal.  Our experiment uses laser
cooled calcium ions confined in a linear radio-frequency
trap.  Measurements for up to 10 ions are in good agreement with
theoretical and numeric predictions.  Implications on an upper limit to the size of data registers
in ion trap quantum computers are discussed.
\end{abstract}

\pacs{PACS numbers:32.80.Pj, 64.60.-i, 03.67.Lx, 52.25.Wz}
\vskip -.2 in





Ions confined in linear radio-frequency traps, and cooled by laser
radiation, will condense into a crystalline state.
Such crystals are the most rarefied form of condensed matter
known \cite{Schiffer}. Besides being of inherent
scientific interest for this reason, cold trapped
ions have a growing number of applications, notably
spectroscopy \cite{Gill,Berkeland,DeVoe},
frequency standards \cite{Berkeland,Fisk}, and quantum computing
\cite{Cirac,sackett}. The existence of  different
kinds of phase transitions of these crystals has been known
for some time \cite{Wineland87,birkl} and has been the subject of various
theoretical and numeric studies \cite{Hasse,Schiffer,Dubin}. Previous experimental work identified different crystal phases/configurations in a quadrupole ring trap \cite {birkl}. Here we explicitly investigate the $\it{transition}$ between two of these phases: the linear and the zigzag configurations.  We report the first experimental confirmation of
one of the key theoretical/numeric predictions for the linear-to-zigzag transition,
namely the existence of a power-law relating the critical anisotropy parameter to the number of ions in the crystal.  Further, we discuss the usefulness of this
power-law expression in determining the ultimate size of a quantum logic register realizable using a single ion trap.

The potential energy of a crystal of $N$
identical ions of mass $M$ and charge $e$
confined in an effective three-dimensional harmonic potential is
\begin{eqnarray}
U\left({\bf r}_{1},{\bf r}_{2},\ldots{\bf r}_{N}\right)&=&
\frac{M (2\pi)^{2}}{2}\sum_{n=1}^{N}\left(
\nu_{x}^{2}x_{n}^{2}+
\nu_{y}^{2}y_{n}^{2}+
\nu_{z}^{2}z_{n}^{2}\right)\nonumber \\
&&+
\frac{e^2}{8\pi\epsilon_0}
\sum_{\stackrel{\scriptstyle n,m=1}{m \neq n}}^N
\frac{1}{|{\bf r}_n-{\bf r}_m|} \ ,
\label{ewe}
\end{eqnarray}
where ${\bf r}_n = \left(x_{n},y_{n},z_{n}\right)$
is the postion vector of the $n-th$ ion, and $\nu_{x}$,
$\nu_{y}$ and $\nu_{z}$ are the trapping potential frequencies in the three directions.  Assume the trapping potentials are approximately
equal in the two transverse directions ($x $ and $y $), so that
$\nu_{x} \approx \nu_{y} \approx \nu_{r}$,
and the trapping potential in the axial ($z$) direction
is different than in the other two directions.
This anisotropy is characterized by the
parameter $\alpha = (\nu_{z}/\nu_{r})^{2}$.
Work by Schiffer \cite{Schiffer} and Dubin
\cite{Dubin} predicts that the ions undergo a phase transition from a linear to
a zigzag configuration at a critical anisotropy value
$\alpha_{crit}$.  The predicted power-law scaling for $\alpha _ {crit} $ versus
the number of trapped ions $N $ is:
\begin{equation}
\alpha_{crit}=c N^{\beta},
\label{scaling}
\end{equation}
where the constants found empirically by Schiffer are $c=2.53$ and $\beta=-1.73$ \cite{misprint}.  Later
in this paper we provide a simple alternative approach which produces qualitative agreement with Schiffer's  results but is based on a stability analysis of the transverse oscillatory modes, rather than on numerical simulations of the equilibrium crystalline configurations.

\begin{figure}[tb]
\begin{center}
\mbox{\epsfig{file=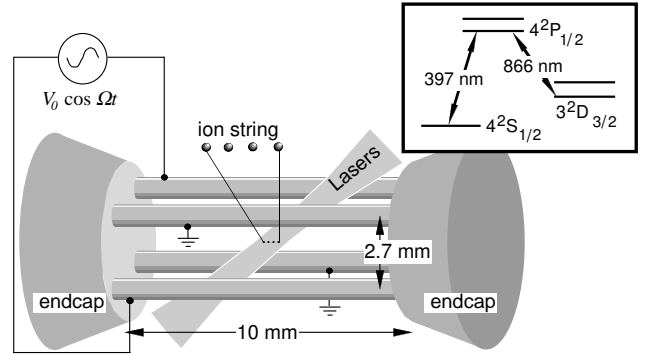,width=3.3in,clip=}}
\end{center}
\caption{ Trap electrodes include four 1-mm diameter rods and two endcaps.  Rf trapping voltages $V_0 $ range from 60-200 V, producing radial frequencies $\nu _r = 200-700 $ kHz.  Endcap voltages range from 10-200 V, producing axial frequencies $\nu _z = 80-390 $ kHz.  Outer support rods (not shown) double as compensation electrodes for moving ions radially within the trap.   The insert shows a level diagram for $^{40}$Ca$^+$.}
\vskip -.15 in
\label{trap}
\end{figure}

We studied the zigzag phase transition using strings of $^{40}$Ca$^+$ ions
confined in a linear radio-frequency quadrupole trap.  Trap electrodes (see
Fig.~\ref{trap} or Ref.~\cite{LANL98}) consist of four rods and two conical
endcaps.  An rf drive at $\Omega/2\pi = 6 $ MHz is applied to two
diagonally opposite rods to produce a radial confining pseudo-potential
\cite{Dawson} with frequency $\nu _ r$ proportional to the rf trapping
voltage $V_0 $ (for example, \cite {Dana}).  Axial confinement is provided
by a static potential applied to the two endcap electrodes, resulting in a
harmonic potential with frequency $\nu _z $ proportional to the square root
of the endcap voltage.  Anharmonic contributions to the axial potential energy are estimated to be less than $3\times 10 ^ {-7} $ the harmonic
contribution, and are thus ignored.

Calcium ions are introduced into the trap by intersecting an
atomic beam with an electron beam.  Laser cooling on the S$_
{1/2} $ to P$_ {1/2} $  transition at 397 nm  (see Fig.~\ref {trap}) cools the ions so that
a crystallized string is formed along the trap axis.  The resulting 397 nm ion
fluorescence is detected in the horizontal plane along a direction
perpendicular to the trap axis by both a CCD camera and a
photomultiplier tube.  An 866 nm laser returns ions falling
into the D$_ {3/2} $ metastable state to the cooling cycle.  The
397 nm light is produced by doubling  794 nm diode-laser light which
is intensified by a tapered amplifier; the 866 nm beam is also
produced by a diode laser.

For each measurement of $\alpha_{crit}$, the rf trapping potential is
lowered, keeping the endcap voltage constant, until the transition to
a zigzag pattern is observed.  The rf potential is then repeatedly raised
and lowered to determine the reproduceability of the transition point and
to rule out hysteresis.  At the identified critical rf voltage, we measure
the radial and axial resonant frequencies.
The measurements are repeated for a range of endcap voltages and then for
different numbers of ions $N$.

To determine the rf trapping voltage at
which the linear crystal becomes zigzag, the voltage is
lowered until the equilibrium position of at least one of the ions
moves visibly (0.5-1.5 pixels $\approx 0.3-1.0 $ $ \mu$m) on the CCD
camera image.  Axial shifts are detected more easily than radial
shifts; therefore, movement in the axial direction is used as the
diagnostic to identify the phase transition.  Moreover, observations show the spacing between the two innermost ions to be
most sensitive to this axial reorganization.  The
critical rf voltage is determined to within $\pm$0.1 V out of
3-8 V on the synthesizer generating the rf trapping voltage.  This
0.1 V resolution, corresponding to 8-16 kHz uncertainty in radial
frequency, is limited by the voltage change necessary to move an ion
visibly on the CCD camera rather than by the synthesizer itself.

The radial and axial frequencies are individually measured by
applying an external drive and observing melting of the crystal into a diffuse but still stable cloud when
the drive frequency is resonant.  Note that although the ions are
driven at a center-of-mass frequency, coupling to higher order modes heats the string until it melts.  The external drive is
the output of a function generator, attenuated and capacitively
coupled through 1000 pF to one endcap electrode.  Resonant
frequencies are measured reproduceably to within $\pm$1-2 kHz by stepping repeatedly back and forth through
the typical 1-4 kHz range in which melting is observed.  Since the
azimuthal symmetry of the trap is not perfect, we measure two
separate radial frequencies, differing from each other by 1-2$\%$.
The data presented includes only the smaller radial frequency
(corresponding to the weaker axis of the radial well) which is
responsible for the zigzag instability onset.

\begin{figure}[tb]
\begin{center}
\mbox{\epsfig{file=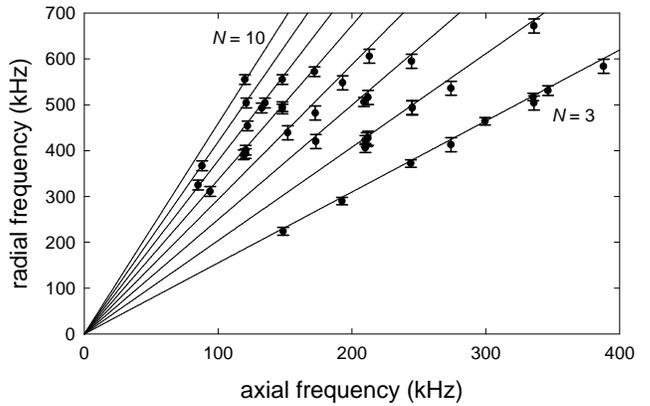,width=3.3in,clip=}}
\end{center}
\caption{Measured radial versus axial frequencies (points), at the onset of the zigzag instability, agree well with the prediction of our theoretical analysis (lines).  Measurements were taken on seven different days, with some days dedicated to a particular length ion string and other days spent studying up to six different length strings.  Theory lines pass through the origin and have slopes $\nu_r/\nu_z=(\alpha_{crit})^{-1/2}$, increasing with ion number for $N = 3 $ (minimum slope shown) through $N = 10 $ (maximum slope shown).  Error bars (see text) are dominated by the uncertainty in determining the critical rf voltage for zigzag onset while the axial frequency is held fixed.    }
\label{rawdata}
\vskip -.15 in
\end{figure}

\begin{figure}[tb]
\begin{center}
\vskip -.2 in
\mbox{\epsfig{file=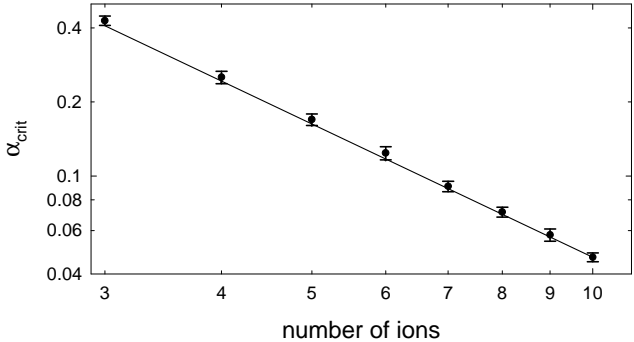,width=3.3in,clip=}}
\end{center}
\caption{Measured values (points) of the critical anisotropy parameter $\alpha_{crit}$ versus ion number $N $ agree well with a fit to the predictions of our theoretical analysis (line).  Each point is a weighted average of measured values from Fig.~\ref {rawdata}, as is each error bar.}
\label{alpha}
\end{figure}

Figure~\ref{rawdata} shows the radial versus axial frequencies at
which the linear-to-zigzag phase transition was observed for string lengths up
to $N = 10 $.  Figure~\ref{alpha} shows the average $\alpha _ {crit}
$  as a function of $N$, and demonstrates the predicted scaling behavior for these critical parameters.
In both figures, the solid lines are results from our theoretical analysis (presented below) with no free parameters.
Although the measurements of $\alpha_{crit}$ lie slightly above
theory, the overall agreement within the 5-6$\% $ error bars is quite
remarkable for a second order phase transition.

In Fig.~\ref {rawdata}, vertical error bars are dominated by the 8-16
kHz resolution in determining the onset of the zigzag mode as described above, but
also include (in quadrature) the 1-2 kHz measurement resolution of
the radial frequency.  Horizontal errors representing the axial
frequency measurement resolution are smaller than the point size and
not shown, but are included when calculating the error for a measured
$\alpha_{crit}$. In Fig.~\ref {alpha}, each plotted value is a
weighted average of measured $\alpha_{crit}$'s.  Corresponding error
bars are taken to be weighted average errors rather than quadrature sums of
the individual $\alpha_{crit}$ errors, because they represent detection
resolution rather than statistical errors and performing many
measurements would not be expected to reduce their size.

\begin{figure}[b!]
\begin{center}
\vskip -.2 in
\mbox{\epsfig{file=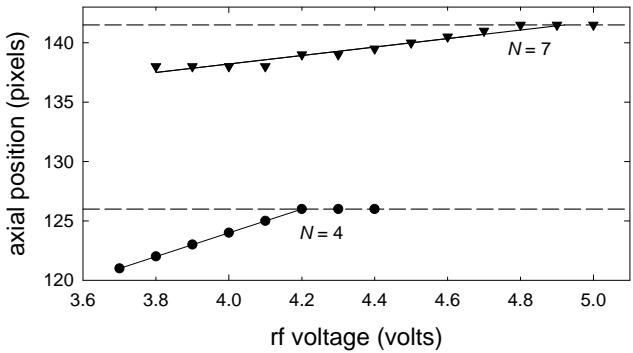,width=3.3in,clip=,}}
\end{center}
\caption{Axial position as a function of rf voltage for an ion nearest to, but not at the center of, a 4 ion string (circles) and a 7 ion string (triangles).   Dashed lines show the steady state axial position for high rf voltages. Solid lines are fits to the points whose rf voltages lie at or below the critical value.}
\label{systematic}
\end{figure}

An unaccounted source of error arises from the threshold axial shift
of 0.5-1.5 pixels required to detect a phase transition.  Any smaller axial
shifts due to onset of the zigzag mode at higher rf voltages go
undetected.  Investigations tracking the axial position of an ion
near the center of a linear crystal as a function of rf trapping
voltage were performed to estimate the size of this detection
threshold error.  Examples such as those shown in
Fig.~\ref{systematic} reveal that once a phase transition is
detected, the axial equilibrium moves roughly linearly with rf
voltage at a rate of 0.3-1.1 pixels per 0.1 V step of the drive
synthesizer.  Not surprisingly, the weaker trap settings -- low
endcap voltage and small ion number --  produce the larger amount of movement per voltage
step but also require the larger threshold
shift.  Underestimates in the critical trapping voltage due to this
effect, therefore, end up of the same order as the 0.1 V
resolution error already assigned.  This detection threshold error
accounts in part for the slightly high trend exhibited by the
measured $\alpha_{crit}$'s.  Alternatively, ignoring the detection
threshold error, our measurement can instead be interpreted as an
upper bound on $\alpha_ {crit}$.

Trapping and cooling parameters were varied outside their normal
ranges to rule out systematic effects.  Detuning the 397 nm laser as
far as possible to longer or shorter wavelengths, while still
maintaining crystallization, altered neither the measured resonant
frequencies nor the measured critical rf voltage.  Similar tests
involving the 866 nm laser wavelength showed no effects.  Finally,
for each new endcap voltage, care was taken to move the ion string to
the radial center of the rf quadrupole potential.  Ions not centered
in this potential experience 
micromotion at the 6 MHz trap frequency.  This driven motion,
detected as a modulation on the ion fluorescence intensity, was used
as a diagnostic to find the optimal position along one radial
direction \cite{Dana}.  The diagnostic of micromotion for the
orthogonal radial direction was the observation of a radial shift in
ion position upon weakening or strengthening the pseudo-potential
well, because an ion's position in the pseudo-potential will depend on the
potential well strength unless the ion is centered \cite{Dana}. Ions were
moved radially by applying voltages to the compensation electrodes, which were typically tuned to $\pm $1 V; tests
detuning them by 5-10 V revealed no change in measured resonant
frequencies or in critical rf voltages.

We now present a simple theoretical analysis of the onset of zigzag
instability.  The equilibrium
positions of the ions are determined by the condition
that the potential be a minimum, i.e.
$\nabla_{n}U=0\,(n=1,\ldots N)$.  For the
case of strong anisotropy ($\alpha \ll 1$), the ions
are configured along a line in the $z$-direction.
The solutions to
the equilibrium equations in this case have been investigated by
various authors (see, for example \cite{Steane,dfvj,todd});
we denote the equilibrium position of the $n-th$ ion by
${\bar{\bf r}}_n = \left(0,0,{\bar z}_{n}\right)$.

Small oscillations of the ions about their equilibrium
positions are described by the Lagrangian \cite
{dfvj,kielpinski}
\begin{eqnarray}
L&=&\frac{M}{2}\left[\sum_{n=1}^{N}\dot{\zeta}^{2}_{n}
-\omega^{2}_{z}\sum_{n,m=1}^{N}A_{n,m}\zeta_{n}\zeta_{m}\right]\nonumber
\\
&+&\frac{M}{2}\left[\sum_{n=1}^{N}\dot{\xi}^{2}_{n}
-\omega^{2}_{x}\sum_{n,m=1}^{N}B_{n,m}\xi_{n}\xi_{m}\right]\nonumber
\\
&+&\frac{M}{2}\left[\sum_{n=1}^{N}\dot{\eta}^{2}_{n}
-\omega^{2}_{y}\sum_{n,m=1}^{N}B_{n,m}\eta_{n}\eta_{m}\right]\ ,
\label{lagrangian}
\end{eqnarray}
where $(\xi_{n},\eta_{n},\zeta_{n})$ are the displacements
of the $n-th$ ion from its equilibrium position in the
$(x, y, z)$ directions, respectively. The coupling matrices
$A_{n,m}$ and $B_{n,m}$ are given by the formulae (following
Ref.~\cite{dfvj})
\begin{eqnarray}
A_{n,m}&=&\left\{
\begin{array}{ll}
1+2{\displaystyle \sum^N_{\stackrel{\scriptstyle p=1}{p \neq m}}
\frac{\ell^{3}}{\left|{\bar z}_m-{\bar z}_p\right|^3}}& {\rm if}\quad
n=m, \\
\rule{0in}{5ex}
{\displaystyle \frac{-2 \ell^{3}}{\left|{\bar z}_m-{\bar
z}_n\right|^3}} &
{\rm if}\quad n\neq m.
\end{array} \right.\nonumber\\
B_{n,m}&=&\left(\frac{1}{\alpha}+\frac{1}{2}
\right)\delta_{n,m}-\frac{1}{2}A_{n,m} \ ,
\label{CoupMats}
\end{eqnarray}
where $\ell=\left[e^{2}/4\pi\epsilon_{0}M (2\pi\nu _z) ^{2}\right] ^{1/3}$
is a length scale and $\delta_{n,m}$ is the Kronecker
delta.

The eigenvectors and eigenvalues of the real, symmetric,
positive-definite matrix $A_{n,m}$ define the normal modes of
oscillation of the ions along the $z$-direction.  Because of their
importance to quantum computing, they have been studied in some
detail \cite{dfvj,kielpinski}. The eigenvectors are defined by the
formula
$\sum_{m=1}^{N}A_{n,m}b^{(p)}_{m} = \mu_{p}b^{(p)}_{n} $,
where
$\mu_{p}>0$ is the eigenvalue, $b^{(p)}_{n}$ the
normalized eigenvector, and
$p$ $(=1, \ldots N)$ the mode index (the modes
being enumerated in order of increasing eigenvalue).

From the definition of the coupling matrix for the
radial oscillations, $B_{n,m}$, we see that it
has identical eigenvectors to $A_{n,m}$, but
has different eigenvalues:
\begin{equation}
\sum_{m=1}^{N}B_{n,m}b^{(p)}_{m} =
\left(\frac{1}{\alpha}+\frac{1}{2}-\frac{\mu_{p}}{2}\right)
b^{(p)}_{n}.
\end{equation}
For these oscillations, the eigenvalues are no longer
always positive.  Indeed, as $\alpha$ is increased,
a critical value occurs for which one of the eigenvalues of the
radial oscillations becomes zero.  Beyond this point the radial
oscillation is unstable and so is the linear configuration: this
marks the onset of the zigzag mode.  The
critical value of $\alpha$ for which this occurs is
given by the exact result $\alpha_{crit}(N)=2/(\mu_{N}-1)$,
where $\mu_{N}$ is the  largest
eigenvalue of the matrix $A_{n,m}$ for a given
number of ions $N$.  The value of $\mu_{N}$
in general must be determined by numerically diagonalizing the matrix $A_{n,m}$ for each value of $N $, but the expression for $\alpha_{crit}(N) $ is exact.  

We can approximate $\alpha_{crit}(N) $ as the power-law in Eq.~\ref{scaling} in order to compare it to our experimental measurements. Table I compares the constants $c$ and $\beta$ (see Eq.~\ref{scaling})
derived from the measurements, from a fit to our theory (over $N = 3-10 $), and from  Schiffer's fit. We emphasize that no matter how well any model
calculates $\alpha _ {crit} $ for a specific $N $, the power-law deduced for
$\alpha _ {crit}$ versus $N$ is still approximate and must be obtained from
a fit of many calculated $\alpha _ {crit} $'s. In light of this, the three
sets of coefficients are close, particularly for $\beta$.  Small
discrepancies may be explained by the fact that the exact expression for $\alpha_{crit}(N) $ is $\it {not} $ a power-law, and that 
Schiffer's predicted values were deduced from a fit of  10 $\alpha_{crit}$
values over the range $N = 2-500 $ whereas we fit over just the
experimental range $N = 3-10$.  Fitting, instead, over all values in the range $N =2-100 $, yields the constants $c = 2.88\pm 0.03 $ and $\beta = -
1.773\pm 0.003$, which are more appropriate (as are Schiffer's) for large $N$ applications of
the power-law discussed below.

\begin{table}[tb]
\label{t}
\caption{Scaling constants $c$ and $\beta$ from fits of the
experimental and theoretical data shown in Fig.~\ref{alpha} and from
Schiffer's numerical results.  Standard errors from our linear
regressions are included.  (The error in Schiffer's linear regression was not reported). An additional $- 5\% $ experimental uncertainty in $\alpha_{crit} $  is included to account for the detection threshold error discussed earlier; it dominates the $- 0.2 $ uncertainty in $c $ and the $\pm 0.04 $ uncertainty in $\beta $.}
\begin{center}
\begin{tabular} {l|ccc}
 & Experiment & Our Theory& Schiffer's Results  \\  
\hline
$c $      & $3.23 ^ {+ 0.06}_{- 0.2} $  &  $2.94\pm 0.07 $ & 2.53\\
$\beta $& $-1.83\pm 0.04 $  & $ -1.80 \pm 0.01 $ & -1.73 \\
\end{tabular}
\end{center}
\vskip -.3 in
\end{table}

Using the power-law expression, we illustrate how to estimate the maximum number of ions it
is reasonable to confine along the trap axis.  This gives one possible upper limit to the  size of data registers for
quantum computation with on-axis ions in a linear trap.  Rewriting
Eq.~\ref{scaling} gives an upper limit $N_{crit}$ to the number of ions one can trap in the linear configuration:  $N_{crit}=(\nu_z/\sqrt{c}\nu_r) ^{2/\beta}$.  Experimental limitations will dictate how large the ratio $\nu_z/\nu_r $ can be made, but this expression should prove
useful in optimizing future designs of ion traps.

In summary, a simple theoretical analysis has produced the
critical anisotropy parameter $\alpha_{crit} $ for the
linear-to-zigzag phase transition.  Experimental measurements have
provided a confirmation of the predicted power-law scaling for $\alpha_{crit} $ versus $N $ which had
until now been untested.  From this scaling we obtain an expression for an upper limit to the size of linear data registers for quantum computation in these traps.

We are grateful to D. J. Berkeland and A. G. Petschek for helpful
conversations. This work was supported by the National Security Agency.





\vskip -.3 in


\begin{references}
\bibitem{Schiffer}
J.~P. Schiffer, Phys. Rev. Lett. {\bf 70},  818  (1993).

\bibitem{Gill}
M. Roberts {\it et~al.}, Phys. Rev. Lett. {\bf 78},  1876  (1997).

\bibitem{Berkeland}
D.~J. Berkeland {\it et~al.}, Phys. Rev. Lett. {\bf 80},  2089  (1998).

\bibitem{DeVoe}
R.~G. DeVoe and R.~G. Brewer, Phys. Rev. Lett. {\bf 76},  2049  (1996).

\bibitem{Fisk}
P.~T.~H. Fisk, Rep. Prog. Phys. {\bf 60},  761  (1997).

\bibitem{Cirac}
J.~I. Cirac and P. Zoller, Phys. Rev. Lett. {\bf 74},  4091  (1995).

\bibitem{sackett}
C.~A. Sackett {\it et~al.}, Nature {\bf 404},  256  (2000).

\bibitem{Wineland87}
D.~J. Wineland {\it et~al.}, Phys. Rev. Lett. {\bf 59},  2935  (1987).

\bibitem{birkl}
G. Birkl, S. Kassner, and H. Walther, Nature {\bf 357},  310  (1992).

\bibitem{Hasse}
R.~W. Hasse and J.~P. Schiffer, Ann. Phys. {\bf 203},  419  (1990).

\bibitem{Dubin}
D.~H.~E. Dubin, Phys. Rev. Lett. {\bf 71},  2753  (1993).

\bibitem{misprint}
The printed values of the parameter $c$ in \cite{Schiffer} appear to be
  misprinted as the reciprocals of their true values.

\bibitem{LANL98}
R.~J. Hughes {\it et~al.}, Fortschr. Phys. {\bf 46},  329  (1998).

\bibitem{Dawson}
P.~H. Dawson, {\em Quadrupole Mass Spectrometry and Its Applications} (Elsevier
  Scientific Publishing Company, New York, 1976).

\bibitem{Dana}
D.~J. Berkeland {\it et~al.}, J. Appl. Phys. {\bf 83},  5025  (1998).

\bibitem{Steane}
A. Steane, Appl. Phys. B {\bf 64},  623  (1997).

\bibitem{dfvj}
D.~F.~V. James, Appl. Phys. B {\bf 66},  181  (1998).

\bibitem{todd}
T.~P. Meyrath and D.~F.~V. James, Phys. Lett. A {\bf 240},  37  (1998).

\bibitem{kielpinski}
D. Kielpinski {\it et~al.}, Phys. Rev. A {\bf 61},  032310  (2000).

\end{references}
\end{document}